\documentclass[twocolumn,showpacs,preprintnumbers]{revtex4}

\usepackage{graphicx}
\usepackage{dcolumn}
\usepackage{bm}
\usepackage{color}
\usepackage{amsmath}
\usepackage{amsfonts}
\usepackage[pdftex]{epsfig}
\usepackage{epstopdf}

\begin{document}

\title{The energy gap of the compound $FeSe_{0.5}Te_{0.5}$ determined by specific heat and point contact spectroscopy.}
\author{Roberto Escudero}
\affiliation{Instituto de Investigaciones en Materiales, Universidad Nacional Aut\'{o}noma de M\'{e}xico. A. Postal 70-360. M\'{e}xico, D.F., 04510 M\'EXICO.}
\author{Rodolfo E. L\'{o}pez-Romero}
\affiliation{Instituto de Investigaciones en Materiales, Universidad Nacional Aut\'{o}noma de M\'{e}xico. A. Postal 70-360. M\'{e}xico, D.F., 04510 M\'EXICO}
\email[Author to whom correspondence should be addressed. Phone: 55526224625 RE, email address:] {escu@unam.mx}

\date{\today}

\begin{abstract}
The superconductor $FeSe_{0.5}Te_{0.5}$ was  studied with Point Contact spectroscopy and specific heat  in  polycrystalline samples. The transition temperature determined by magnetic measurement was  $T_C =14.5$ K. The size of the energy gap measured  by junctions  is  $\Delta = 1.9$ meV, whereas the gap determined by the specific heat measurements was $\Delta = 2.3 $ meV. The gap evolution with temperature follows BCS, the  ratio  2$\Delta$/$K_BT_C$ has  values between 2.88 $\leq  2\Delta/K_BT_C \leq 3.04$.
The compound was  grown by solid state synthesis in quartz ampoules under  vacuum at   950 C. Crystal structure was characterized by X-ray diffraction. The superconducting properties were characterized  by magnetization, resistivity and specific heat. This superconductor shows an isotropic energy gap as observed with the  fitting of the specific heat at low temperature.
\end{abstract}

\maketitle
\section{Introduction}
The  new superconducting materials FeSe is interesting  because they may give guides  to elucidate the  superconducting mechanism of  other similar  members with Fe and to others of the  FeAs family. FeSe compound is one of the simplest Fe-Based superconductor  with a  transition temperature  $T_C \simeq  8$ K.  It has    crystalline  structure  type anti-PbO  and space group P4/nmm \cite{hsu,mizu,Lehman1}.

Compounds in these  families  have different transition temperatures,  some with higher transition temperatures as  the basic FeSe \cite{Yoichi, kami,mizu,mcqueen,mizu2}. Recently many study have been performed on these  compounds \cite{mizu,kami, mizu2,singh, sales} some with  transition temperatures as high as $T_C = 40 K $.

In those compounds the magnetism of Fe may play a relevant role which still is not totally understood for the superconducting behavior.

In this work we report a study of the superconductivity in Fe-Se-Te performed in polycrystalline samples. The study was mainly directed to observe and analyze the behavior of the energy gap and to determine the symmetry of the pairing wave function. In general  for the determination of the energy gap, two tools that are well appropriate; Point Contact Spectroscopy (PCS) and Tunnel junctions. We used in this study one of those spectroscopic techniques. The junctions  were  formed with  the superconducting compound with stoichiometry $FeSe_{0.5}Te_{0.5}$ and a normal metal. The sample has a high transition temperature about $ T_C = 14.5$ K, as determined by Magnetic - Temperature measurements. Our studies show the size of the energy gap, the ratio $2\Delta/K_BT_C $, and the evolution with temperature of the gap. In addition, specific heat measurements indicated an isotropic energy gap with $s$ symmetry, the superconducting gap follows the BCS theory. The junctions were characterizes with the BTK model \cite{btk}. With this model we can  describe the type of junctions; i.e. PCS or tunnel, using the dimensionless barrier strength parameter $Z$ given by the theory. Depending of this parameter the junctions may be considered tunnel or as a metallic contact (PCS).

\section{Experimental Details}

\begin{figure}
\centerline{\includegraphics[scale=0.63]{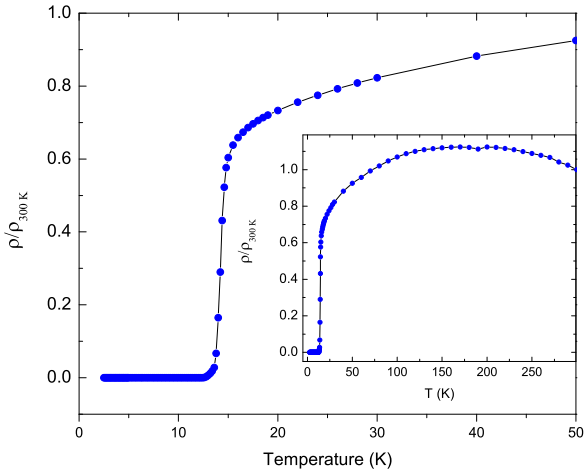}} \caption{(Color online) Resistivity - Temperature characteristic of the superconducting compound $FeSe_{0.5}Te_{0.5}$. Main panel displays the temperature interval close to the transition temperature, $T_C = 14.5$ K. Inset shows the overall characteristics.}
\label{fig1}
\end{figure}

\subsection{compound and junctions characteristics}

The superconducting samples were prepared by solid state reaction in evacuated quartz tubes with powder and purities; Fe 99.9\%, Se 99.9\%, and Te 99.99\%. The nominal proportion was 1.01:0.5:0.05, the powders were mixed pressed and sintered at 950$^\circ$C for three days. Finally, annealed at 400 - 500$^\circ$C in a period of one day. R-X diffraction patterns gave the crystalline structure as the reported by \cite{cao, Awana,sales, Migita, Das, JorgeLuiz, Despina, Tsurkan}.

The study was performed with junctions that behave as metallic point contacts PCS.  The junctions were prepared with the superconductor and a thin wire of tungsten plated gold W(Au) which has a diameter of 5 $\mu$m, the stoichiometry of the superconductor was $FeSe_{0.5}Te_{0.5}$. For the fabrication of the PCS the wire was diagonally clipped as the procedure used in tips for use in scanning tunneling microscopy and carefully pressed into the superconducting sample. This cut allows to obtain very sharp tips with very small diameters about or less that 1 $\mu$m \cite{Bai}.
Oxford varnish was used to glue the sample to a glass support. The estimated area of the junctions was $\sim$ 1 $\mu$m$^2$, as observed in an optical microscopy.  The determined parameters with the BTK model gave values for the $Z$ parameter from 3.4 to 6 \cite{btk}. This was one of the main considerations to determinate the type of junction.

 At the initial measurements several try outs were performed until reproducible data and stable characteristics were obtaining. Results presented in this work are the most reproducible for all the junctions.

Many junctions, more that 25 of PCS's  were prepared. The characteristics displayed in figures 4,5,  were the most reproducible. The characteristics were measured from 1.7 to 25 K. Results show the size of the energy gap, the gap evolution with temperature, and ratio 2$\Delta/K_BT_C$. The compound characteristics were determined by resistivity, $\rho$ - T, specific heat, $C_P - T$, and magnetization, $M-T$, accordingly to transport and magnetic measurements the transition temperature was $T_C =14.5$ K, as shown by other researchers \cite{mcqueen,mizu2,cao,Migita}.

\begin{figure}
\centerline{\includegraphics[scale=0.73]{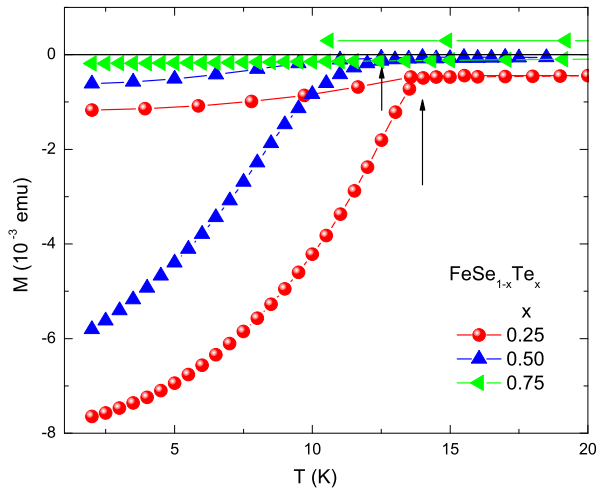}}
\caption{(Color online) Magnetization measurements of three different compound with different stoichiometries. The compound  $FeSe_{0.5}Te_{0.5}$,  was  selected for the study. This shown  the maximum proportion of superconducting material. The  panel displays the $M-T$ measurements in the temperature interval close to the transition temperature in  ZFC and FC modes in order to determinate  the amount of superconducting material. The transition temperature was $T_C = 14.5$ K.}
\label{fig2}
\end{figure}

The superconducting characteristics were determined in a Quantum Design MPMS system, the amount of superconducting phase was estimated by Zero Field Cooling (ZFC) and Field Cooling (FC) measurements  with 10 Oe. The  proportion of superconducting material was $\sim$ 70\%, compared with pure Pb with the same mass as  the sample. Transport characteristic were determined in  a Quantum Design PPMS system,   Fig. 1  displays  the resistivity versus temperature for the selected sample. Main panel shows the  resistivity $\rho-T$ close to the transition temperature and  the inset presents the overall behavior to high temperature. In Fig. 2 we display the magnetic  characteristics of the three different compositions. The sample with the composition  $FeSe_{0.5}Te_{0.5}$, was the used for  determination of the spectroscopic characteristics  because it  has the maximum proportion of superconducting material.

\begin{figure}
\centerline{\includegraphics[scale=0.7]{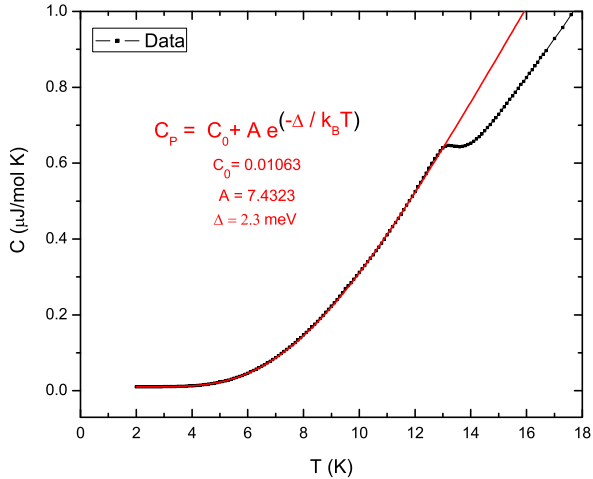}} \caption
{(Color online) Specific heat of the superconducting compound $FeSe_{0.5}Te_{0.5}$. Main panel displays  the fitting of $C_P$ experimental data to the function $C_P = C_0 + Ae ^{(-\Delta/K_BT)}$, from 14 to 2 K (red line). According to the specific heat measurements the transition temperature is about $T_C =13.3$ $\pm 0.03$ K.}
\label{fig3}
\end{figure}

  The isotropic characteristics of the energy gap were determined by thermal characterization measuring the  specific heat capacity. The main panel in Fig. 3 displays the observed characteristic close to the transition temperature. The specific heat measurements, $C_P - T$, were performed with a thermal relaxation method using a Quantum Design PPMS apparatus. The $C_P$ values were corrected by subtracting the addenda of the sample support and grease used to glue the sample to the support. Thermal measurements show similar overall characteristics as determined by Sales, and Awana. The Debye temperature was $\Theta_D = 171-174 K$ \cite{sales, Awana}. At low temperature specific heat measurements may give indications related to the type of superconducting pairing \cite{Dong}. Our observations at low temperature, show a decreasing that fit quite well to an exponential decay, as $C_P = C_0 + Ae ^{(-\Delta/K_BT)}$. Two different parameters for the fitting were used as shown in the main panel of Fig. 3, $C_0$ and $A$ are parameters to adjust the background level of the specific heat curve at low temperature, whereas $\Delta = 2.3$ meV. The important result is that the superconducting compound is isotropic with a single gap and $s-wave$ symmetry, the size of the energy gap is similar to as determined by the PCS, and the transition temperature, was about $T= 13.3 \pm 0.03$ K. However, it is important to mention that determination of the energy gap with specific heat measurements is not as precise as tunneling or PCS measurements, nevertheless, specific heat may give clear information about features of multi-gap or single gap characteristics.
\begin{figure}
\centerline{\includegraphics[scale=0.7]{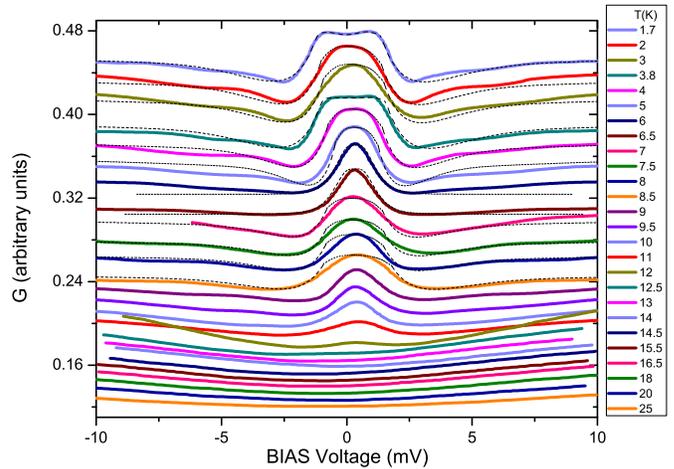}}
\caption{(Color online) Differential conductance of a contact  formed with a  W(Au) wire tip  and the  compound $FeSe_{0.5}Te_{0.5}$. Figure shows the differential conductance at different temperatures in the region close to the features of the  energy gap. All curves  were displaced vertically by a small amount of the differential conductance  for a better view of the general behavior. Values extracted with the BTK model give $Z$ = 3.42, and $\Delta$ = 1.9 meV. The fitting $Z$ were taken into the same context as to the BTK model. This  was performed  for  data taken from 1.7 to 8.5 K as shown in  figure with black pointed lines. The fitting is clearly seen  in Fig. 5 plotted  in one curve of the normalized differential conductance at a temperature  of $T=1.7 K$. }
\label{fig4}
\end{figure}

\begin{figure}
\centerline{\includegraphics[scale=0.7]{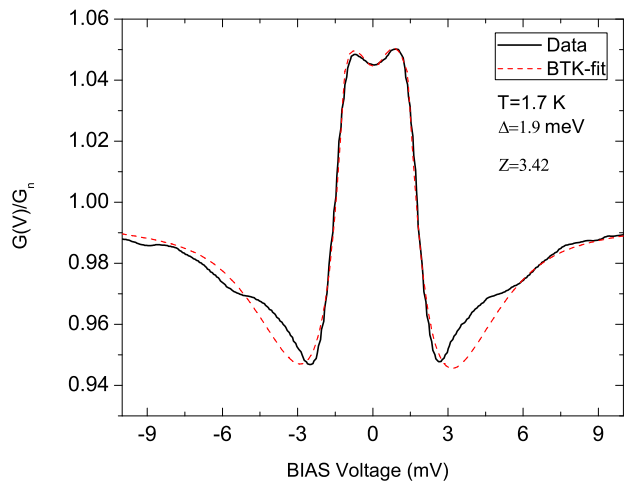}}
\caption{(Color online) Normalized differential conductance for the PCS data of Fig. 4.  The figure displays data measured  at  $T= 1.7$ K (black lines). Parameters determined with the BTK model (dotted red line) give values for the energy gap, $\Delta = 1.9$  meV,  $Z = 3.42$, and  ratio 2$\Delta$/$K_BT_C$, as 2.88 $\leq  2\Delta/K_BT_C \leq 3.04$.}
\label{fig5}
\end{figure}

For the point contact measurements the differential resistance d$V$/d$I$ as a function of the bias voltage
$V$ for all junctions was measured with the standard techniques; AC Lock-in amplifier, and bridge \cite{wolf,Adler}. A Quantum Design MPMS system was used as a cryostat.
The junctions were measured in the temperatures range from 1.7 to 25 K. PCSs were close to the ballistic regime according to the electronic parameters of the compound and to the values of the differential resistance (at zero bias voltage) and $Z$ parameter. In figures 4 and 5 we show the overall characteristics of the PCS's. Fig. 4 displays the general behavior of the differential conductance versus bias voltage at different temperatures measured from 1.7 to 25 K. In this set of measurements the BTK model fits well to the experimental data at low temperatures from 1.7 to about 7.5 K. With this result, it was possible to extract values for $Z$, which were $Z = 3.42$, and  energy gap $\Delta = 1.9$ meV. The $Z$ parameter used in this study has the same meaning as the mentioned in the BTK theory; this, measure the barrier strength at the interface. As one example for the resulting fitting Fig. 5 shows the extracted parameters. In Fig. 6 we show the evolution of the gap with temperature plotted in terms of $\Delta - T$. The experimental data was plotted only at low temperature region, data at higher temperatures was difficult to fit as mentioned in description of Fig. 4.
The characterization of the work regime of the PCS's  was estimated with the Wexler's interpolation formula \cite{wexler}; we substituted the mean free electronic path estimated as $l\simeq 100$ \AA   using the Drude model and  according to the resistivity measurements \cite{ashcroft},
the resistivity $\rho \sim $ 40 $\mu\Omega $-cm measured at 2 K and the resistance of the junction measured at zero
bias voltage. The obtained radii values, were between 320  $\AA$   and  3700  $\AA$, which indicate
that some contacts are in the diffuse regime, and others in the ballistic limit, only PCS's in the ballistic regime, or close to it were used to determined the spectroscopic features \cite{duif,Jansen,Daghero}.

No broadening parameter was  used for the determination of the spectroscopic characteristics as the one introduced by Dynes et.al., \cite{dynes}.

\begin{figure}
\centerline{\includegraphics[scale=0.7]{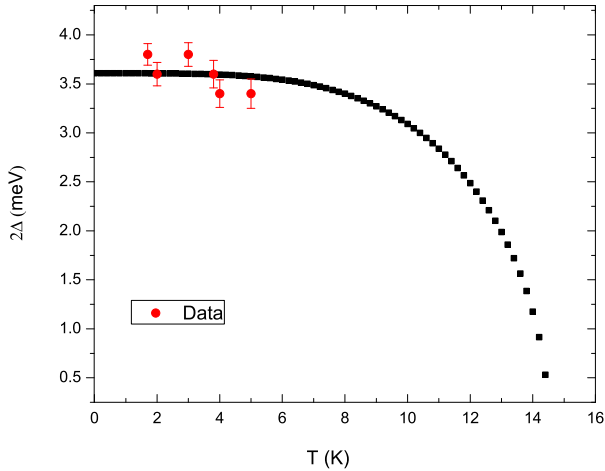}}
\caption{(Color online) Evolution of the energy gap with temperature obtained from  curves of the differential conductance of the PCS's data shown in Fig. 4 (red dots), fitted with BTK model. Black squares are the BCS theory. The fit was performed only in the low temperature range as explained in the main text.}
\label{fig6}
\end{figure}

\section{Results, Discussion and conclusions}

 Junctions with different characteristics behaving as PCS  were  used  to investigate the nature of the superconducting state in a  $FeSe_{0.5}Te_{0.5}$  compound.  PCS and tunneling are tools  well appropriated to study electronic processes occurring close to the Fermi surface,  as  the energy gap and the evolution with temperature and  ratio $2\Delta/K_{B}T_{C}$.
Our measurements with specific heat give us information about the pairing and symmetry state of the compound, and information about the characteristics of the order parameter. Our analyses indicate the characteristics of the pairing symmetry accordingly to the experimental data this superconducting compound  behaves as a full $s - wave$ BCS superconductor. A single feature of the gap was observed in thermal measurements corroborated the shape of the decreasing of the specific heat below $T_C$. An exponential decay of the specific heat with temperature was observed. It shows a good fitting to an exponential rate of decreasing,  indicative that a single gap exist. The decreasing of $C_p -T$ below the transition is quite similar to the observations by Dong, et al., \cite{Dong} in $FeSe_{x}$. In our work the $C_P - T$ curve can be well fitted to the exponential function without signs of a multi-gap system.

It is important to mention that Li, et al., \cite{Can-Li} reported the evidence of nodes in a compound with a similar stoichiometry as in this work. The observation of the gap features by Li, et al., shown a wave pairing different from $s-wave$, that could be $p$, or $d$. However, the multigap feature determined by Li, et al., was measured close to a vortex state and the information could be distorted the real features. Lastly, other important probes appropriated to determine types of Cooper pairs was using ARPES. With this technique as observed by Miao, et al., found that at the composition $FeTe_{0.55}Se_{0.45}$, which is quite similar to our shown only an isotropic single energy gap \cite{miao}.

Our  conclusions of this  work is  that  this superconductor behaves as  a $s-wave$ symmetry with a clear evolution of the energy gap with temperature of the  BCS type. The values for the size of the energy gap with the PCS junctions gave   $\Delta = 1.9$ meV, whereas with  specific heat measurements the energy gap determined was  $2\Delta = 2.3$ meV. Other important parameter of this compound is the ratio $\Delta/K_{B}T_{C}$ = $2.96\pm 0.05$, accordingly, this superconducting material is  in  weak coupling limit.

\begin{acknowledgements}

This work was supported by CONACyT project 129293, DGAPA-UNAM project IN106014, BISNANO, and ICYTDF, project PICCO. RELR was supported by DGAPA-UNAM project. We thanks to J. Morales, A. Lopez,  for  help in computational and technical problems, and to F. Silvar for He provisions.
\end{acknowledgements}

\end{document}